\documentclass[a4paper,10pt,twoside]{cpc-hepnp}
\usepackage{upgreek,fancyhdr}
\usepackage{multicol,color}
\usepackage{graphicx,subfigure,dcolumn}
\usepackage{epsfig}
\usepackage{booktabs,slashed}
\usepackage{amssymb,bm,mathrsfs,bbm,amscd,amsfonts}
\usepackage[tbtags]{amsmath}
\usepackage{lastpage}
\usepackage{multirow}
\usepackage{CJK}

\newcommand{\f}{\begin{equation}}
\newcommand{\ff}{\end{equation}}
\newcommand{\fa}{\begin{eqnarray}}
\newcommand{\ffa}{\end{eqnarray}}

\providecommand{\U}[1]{\protect\rule{.1in}{.1in}}
\begin{document}

\title{\boldmath Entanglement in simple spin networks with a boundary
\thanks{We are very grateful to Yuxuan Liu and Zhuoyu Xian for helpful
discussions and suggestions. This work is supported by the Natural
Science Foundation of China under Grant No. 11575195 and 11875053.
Y.L. also acknowledges the support from Jiangxi young scientists
(JingGang Star) program and 555 talent project of Jiangxi
Province.}}

\author{Yi Ling  $^{1,2}$ \email{lingy@ihep.ac.cn}
\quad Meng-He Wu  $^{1,2}$ \email{mhwu@ihep.ac.cn}
\quad Yikang Xiao  $^{1,2}$ \email{ykxiao@ihep.ac.cn}}

\maketitle

\address{
$^1$ Institute of High Energy Physics, Chinese Academy of Sciences, Beijing 100049, China\ \\
$^2$ School of Physics, University of Chinese Academy of Sciences, Beijing 100049, China}
\maketitle

\begin{abstract}
We investigate the bipartite entanglement for the boundary states
in a simple type of spin networks with dangling edges, in which the
two complementary parts are linked by two or more edges. Firstly,
the spin entanglement is considered in the absence of the
intertwiner entanglement. By virtue of numerical simulations, we
find that the entanglement entropy usually depends on the group
elements. More importantly, when the intertwiner entanglement is
taken into account, we find that it is in general impossible to separate the total entanglement entropy
into the contribution from spins on edges and the contribution
from intertwiners at vertices. These situations are in contrast
to the case when the two vertices are linked by a single edge.

\end{abstract}



\section{Introduction} \label{sec:intro}
Entanglement is the prominent phenomenon in quantum physics.
Recently, it has been discovered that it also plays a key role in
understanding the emergence of spacetime in the framework of
holographic gravity\cite{Maldacena:2001kr}. On the one hand, the
Ryu-Takayanagi (RT) formula provides a geometric description for
the entanglement entropy of a subsystem on a boundary, which is
measured by the area of the minimal surface in the
bulk\cite{Ryu:2006bv}. Such an area law is analogous to the
Bekenstein-Hawking entropy for black holes. On the other hand, the
behavior of quantum entanglement reflects the structure of the
spacetime such that the background information can be extracted
from the correlations of quantum states in a many-body
system\cite{VanRaamsdonk:2010pw}. In particular, it turns out that
the holographic properties of $AdS$ spacetime can be captured by
various types of tensor network states, such as multiscale
entanglement renormalization ansatz
(MERA)\cite{Vidal:2007hda,Swingle:2009bg,Swingle:2012wq,Nozaki:2012zj,Qi:2013caa},
perfect tensor networks\cite{Pastawski:2015qua}, as well as
hyperinvariant tensor
networks\cite{Evenbly:2017htn,Ling:2018vza,Ling:2018ajv}.

Above attempts of investigating the structure of spacetime by
entanglement are background dependent. In particular, the RT
formula is proposed in the large N limit such that the
perturbations in the bulk are controlled by the classical Einstein
equations. It is quite intriguing to explore the role of quantum
entanglement in the emergence of spacetime in a background
independent manner, because the holographic nature of gravity
is believed to be at the core of the quantum theory of gravity, which
is beyond the large N limit of the gauge theory in standard
AdS/CFT correspondence, where the bulk geometry is fixed and
higher order corrections to gravity are greatly suppressed. When
the gravity is strong enough, the dynamics of the bulk geometry
can not be treated in a perturbative manner. One has to face the
quantum nature of the background when building the
geometry of the spacetime from the microscopic point of view by virtue
of entanglement. In loop quantum gravity, it is well known that
the geometry of spacetime itself can be quantized and the quantum
states of the gravitational field are described by spin network states,
which are $SU(2)$ gauge invariant in four dimensional
spacetime\cite{Rovelli:1994ge,Rovelli:1995ac}. Thus spin networks
provide a very  clear description of the atomic structure of the
quantum geometry. In the traditional treatment, spin network
states are mainly considered for closed graphs with fixed spins
and intertwiners, such that they form a set of basis states in the
Hilbert space of the gravitational field. It is clear that for a
closed graph, a spin network is just a basis state without
carrying any entanglement. Thus, in the past the entanglement
structure of spin networks has rarely been addressed.
Recently, the role of entanglement in building
the geometry of spacetime  has been revealed\cite{VanRaamsdonk:2010pw}, and several publications on the relationship between quantum
entanglement and spin networks have appeared\cite{Orus:2014poa,Han:2016xmb,Chirco:2017vhs,Chirco:2017xjb,Livine:2017fgq,Baytas:2018wjd}.
Basically, in the context of spin networks, the possible
entanglement comes in the following two ways: the first is to
consider the superpositions of intertwiners and spins, or many spin
network states, while the second is to consider the spin
networks for an open graph with dangling edges. In
ref. \cite{Han:2016xmb}, the notion of spin networks has been extended
to the non-closed graphs with dangling edges to describe the
quantum geometry with a boundary, and the RT formula is understood in
the coarse graining process. In this context, the $SU(2)$ gauge
invariance is only imposed on the internal vertex, while the
uni-valent vertices linked to dangling edges are not gauge
invariant. The associated degrees of freedom become physical
on the boundary and are described by the boundary spin states.
In ref. \cite{Livine:2017fgq}, the entanglement structure is
investigated for a specific type of spin networks in which two
neighboring vertices are linked by a single edge, and the notion of intertwiner entanglement is proposed.
Moreover, the contribution from intertwiner entanglement at
vertices and spin entanglement on edges are separated.
Interestingly, one finds in this case that the spin entanglement from
the edge, with irreducible representation $j$, always contributes to
the entanglement entropy with the term $\ln(2j+1)$, which is
independent of the group elements.

The separation of spin entanglement and intertwiner
entanglement in a network looks peculiar if one recalls the
nonlinear nature of entanglement entropy. One may speculate if it
is always possible to separate the entropy into these two
contributions in a general spin network. This clarification would
improve our understanding of the structure of entanglement in spin
networks. Therefore, in this paper we further develop the
results of \cite{Livine:2017fgq} by considering a more
situation of spin network with dangling edges where two
neighboring vertices are linked by two or more edges, in
either direct or indirect manner. We investigate the
bipartite entanglement entropy associated with the boundary
degrees of freedom on dangling edges. Moreover, for simplicity, we
perform numerical analysis for a simple type of spin networks
containing two multi-valent vertices or several tri-valent
vertices. We believe that the results are general enough and could be
applicable to more complicated spin networks. We first consider
the spin entanglement from edges in the absence of intertwiner
entanglement. By virtue of numerical evaluations we
demonstrate that, in general, the entanglement entropy depends on
the group elements on edges, which has previously been pointed out
in ref. \cite{Livine:2017fgq}. Our numerical results imply that
once the spins on edges are defined, bounds for the spin
entanglement should exist. Secondly, we consider the
bipartite entanglement entropy in the presence of intertwiner
entanglement. In this case, we find that, in general, it is not
possible to separate the total entropy into spin entanglement and
intertwiner entanglement. Mathematically, it can not be written as
a sum of two distinct parts any more. Our conclusions and
outlook are given in the last section.

\section{Entanglement in the absence of intertwiner entanglement} \label{sec-eae}

In this section, we evaluate the bipartite entanglement
entropy for a few spin networks in the absence of
intertwiner entanglement. First, we consider the case when the
two neighboring vertices are linked by two edges directly. In
general, a spin network is a graph $\Gamma$ composed of edges and
vertices, which could be closed or non-closed. The spin network
state for a non-closed graph $\Gamma$ with dangling edges $o$ is
denoted by $|\Gamma, \{j_e,j_o\},\{I_v\},\{M_o\}\rangle$, where
$j_e$ denotes the spin on the internal edge $e$ and $I_v$ denotes
the intertwiner at internal vertex $v$, while spin
$j_o$ and magnetic quantum number $M_o$ are assigned to each
dangling edge $o$. The corresponding spin network function can
be written as
\begin{eqnarray}\label{eq:eps+2}
\left \langle \{ h_{e}\}, \{ h_{o}\}  |  \Gamma, \{j_{e}, j_{o}
\}, \{ I_{v}  \} ,  \{ M_{o} \}  \right \rangle &=& \sum_{m_e, n_e,
m_o} \prod_{e}  U_{m_e n_e}^{j_e}(h_{e}) \prod_{v} (I_v)_{\{ m_e,
n_e, m_o \}}^{ \{j_e j_o \} } \prod_{o}  U_{m_o M_o}^{j_o}(h_{o}),
\end{eqnarray}
where $h_{e}$ and $h_{o}$ are holonomies along the internal edge
$e$ and dangling edge $o$, respectively, and $U^j$ is the
matrix representation of SU(2) group with spin $j$. This kind of
spin networks is constructed for a spatial region with a boundary.
The total Hilbert space is composed of the
Hilbert space associated with the bulk $\textbf{H}$ and the
Hilbert space associated with the boundary
$\textbf{H}^{\partial}$. Thus, a spin network state can be written
as the direct product of two parts, namely $|\Gamma,
\{j_e,j_o\},\{I_v\},\{M_o\}\rangle = \otimes | I_{v}^{\{j_e,j_o\}
}  \rangle  \otimes  |  j_o, M_o \rangle$.  We define a
boundary state  $|\widetilde{\Psi} [\Gamma, \{j_e, j_o\},\{ I_v
\},\{h_o\}]\rangle \in \textbf{H}^{\partial}$, such that $\otimes
\left \langle j_o, M_o |\widetilde{\Psi} [\Gamma, \{j_e, j_o\},\{
I_v \},\{h_o\}]\right \rangle  = \left \langle \Gamma, \{j_{e},
j_{o}  \}, \{ I_{v}  \} , \{ M_{o} \} |  \{ h_{e}\}, \{ h_{o}\}
\right \rangle$. Next, we study the bipartite entanglement in
the boundary spin state $|\widetilde{\Psi} [\Gamma, \{j_e,
j_o\},\{ I_v \},\{h_o\}]\rangle $.

For simplicity, we first consider a spin network with only
two vertices $A$ and $B$. When the spins on edges are defined,
the Hilbert space of the bulk is given by the products of two
intertwiners,
\begin{eqnarray}\label{eq:eps+3}
\mathbf{H}_{AB}&=& \mathbf{H}_{A}\otimes \mathbf{H}_{B}
\end{eqnarray}
where $ \mathbf{H}_{A}$ and $\mathbf{H}_{B}$ are the spaces of
intertwiners attached to two vertices $A$ and $B$, respectively,
\begin{eqnarray}\label{eq:eps+4}
\mathbf{H}_{A} &=& \textup{Inv}_{SU(2)}[V^{J_1}\otimes ...V^{J_p}\otimes V^{j_{1}} \cdots \otimes V^{j_{n}}] \nonumber \\
\mathbf{H}_{B} &=& \textup{Inv}_{SU(2)}[V^{K_1}\otimes
...V^{K_q} \otimes V^{j_{1}} \cdots \otimes V^{j_{n}}],
\end{eqnarray}
where we have assumed that $p$ dangling edges with spins
$J_p$ are joined to vertex $A$, $q$ dangling edges
with spins $K_q$ are joined to vertex $B$, and the two
vertices are linked directly by $n$ internal edges with spins
$j_n$.

The Hilbert space of the boundary spin states is
\begin{eqnarray}\label{eq:eps+5}
\mathbf{H}^\partial_{AB} &=& \mathbf{H}^\partial_{A}\otimes
\mathbf{H}^\partial_{B},
\end{eqnarray}
where $ \mathbf{H}^\partial_{A}$ and $\mathbf{H}^\partial_{B}$ are
the spaces of spins on dangling edges joined to vertices $A$ and
$B$, respectively,
\begin{eqnarray}\label{eq:eps+6}
\mathbf{H}^\partial_{A} &=& V^{J_1}\otimes V^{J_{2}} \cdots \otimes V^{J_{p}} \nonumber \\
\mathbf{H}^\partial_{B} &=& V^{K_1}\otimes V^{K_{2}}
\cdots \otimes V^{K_{q}}.
\end{eqnarray}

\begin{figure}
  \center{
  \includegraphics[scale=0.4]{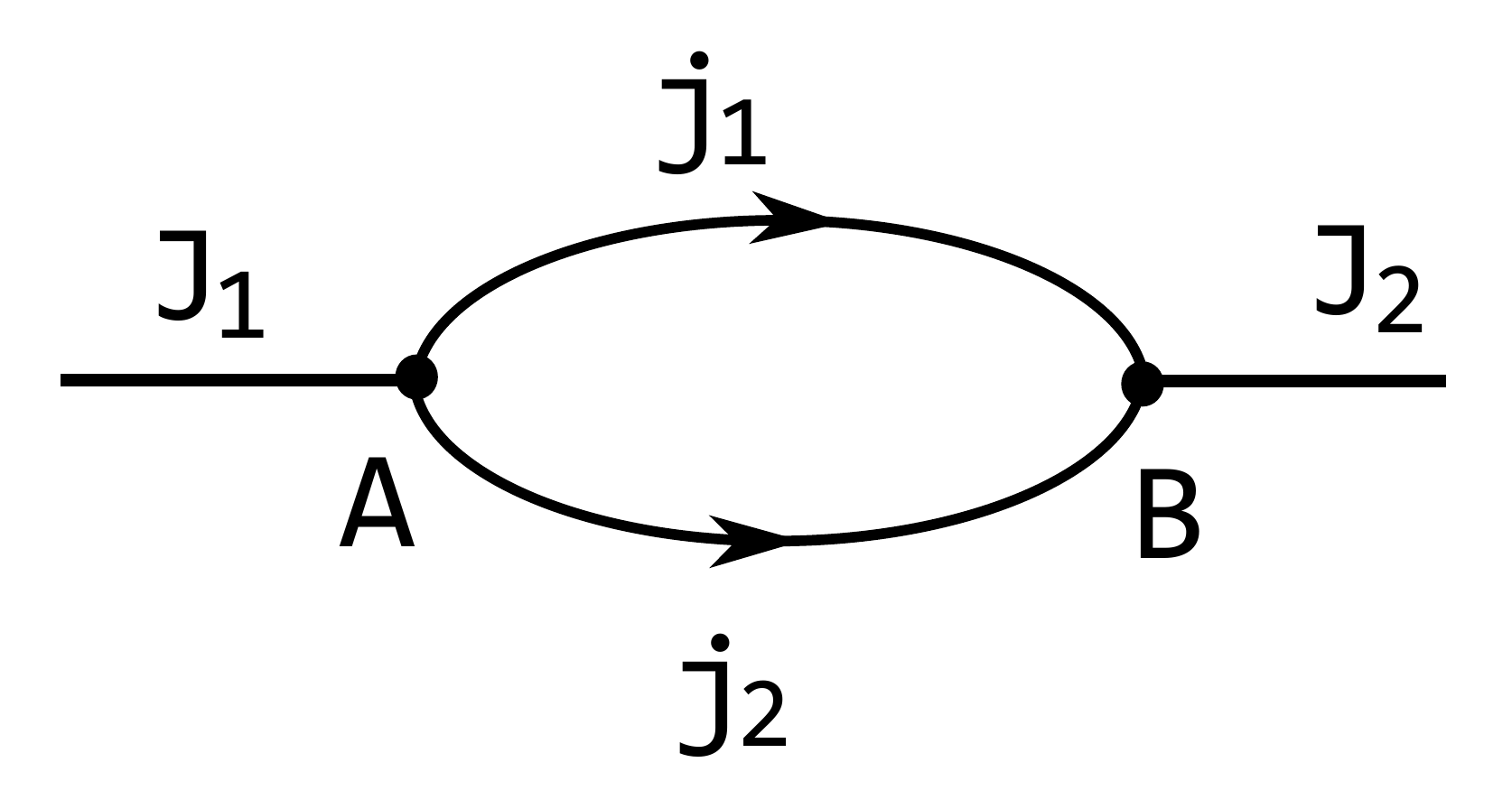}\ \hspace{0.05cm}
  \includegraphics[scale=0.25]{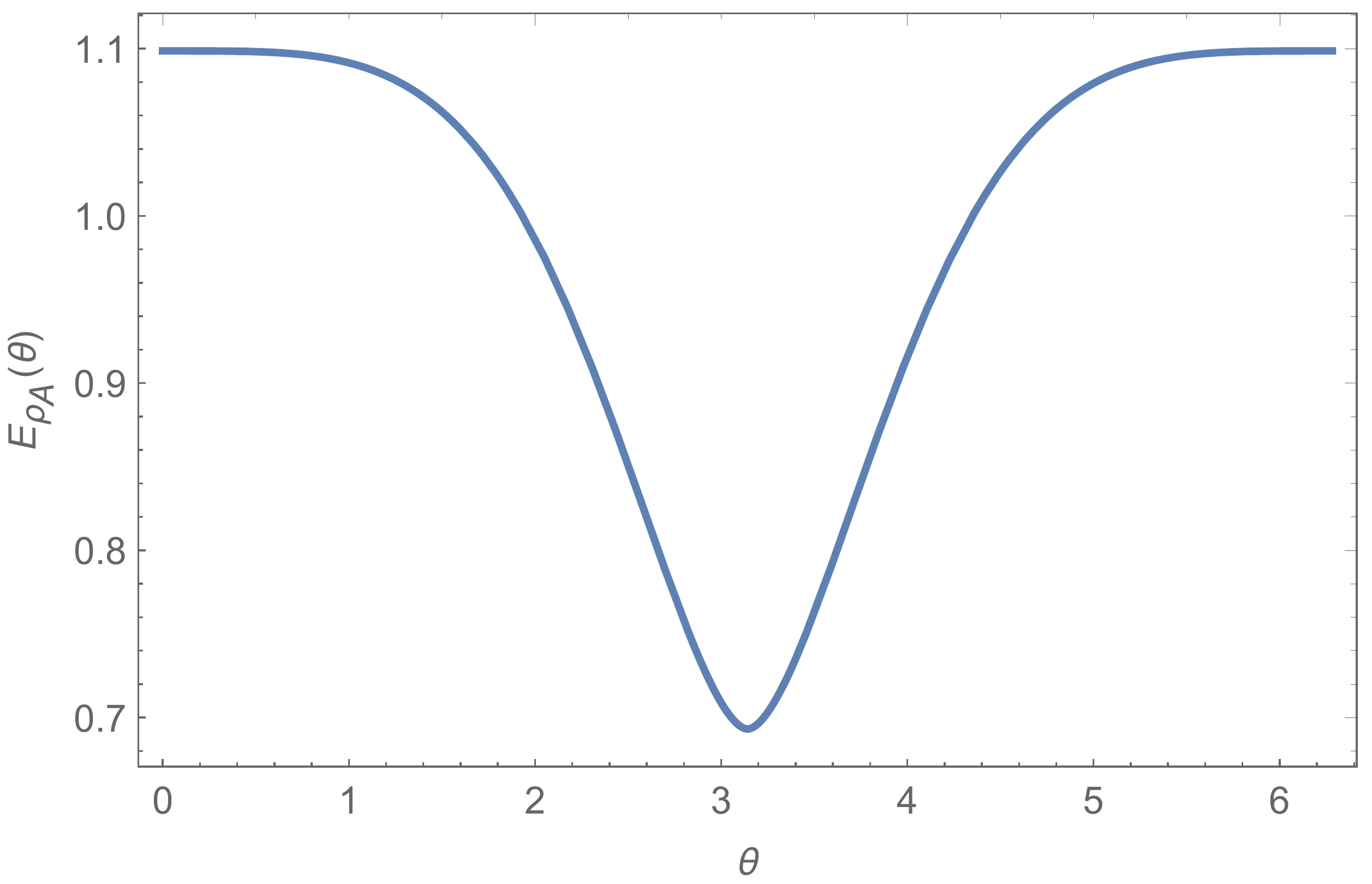}\ \hspace{0.05cm}
  \includegraphics[scale=0.26]{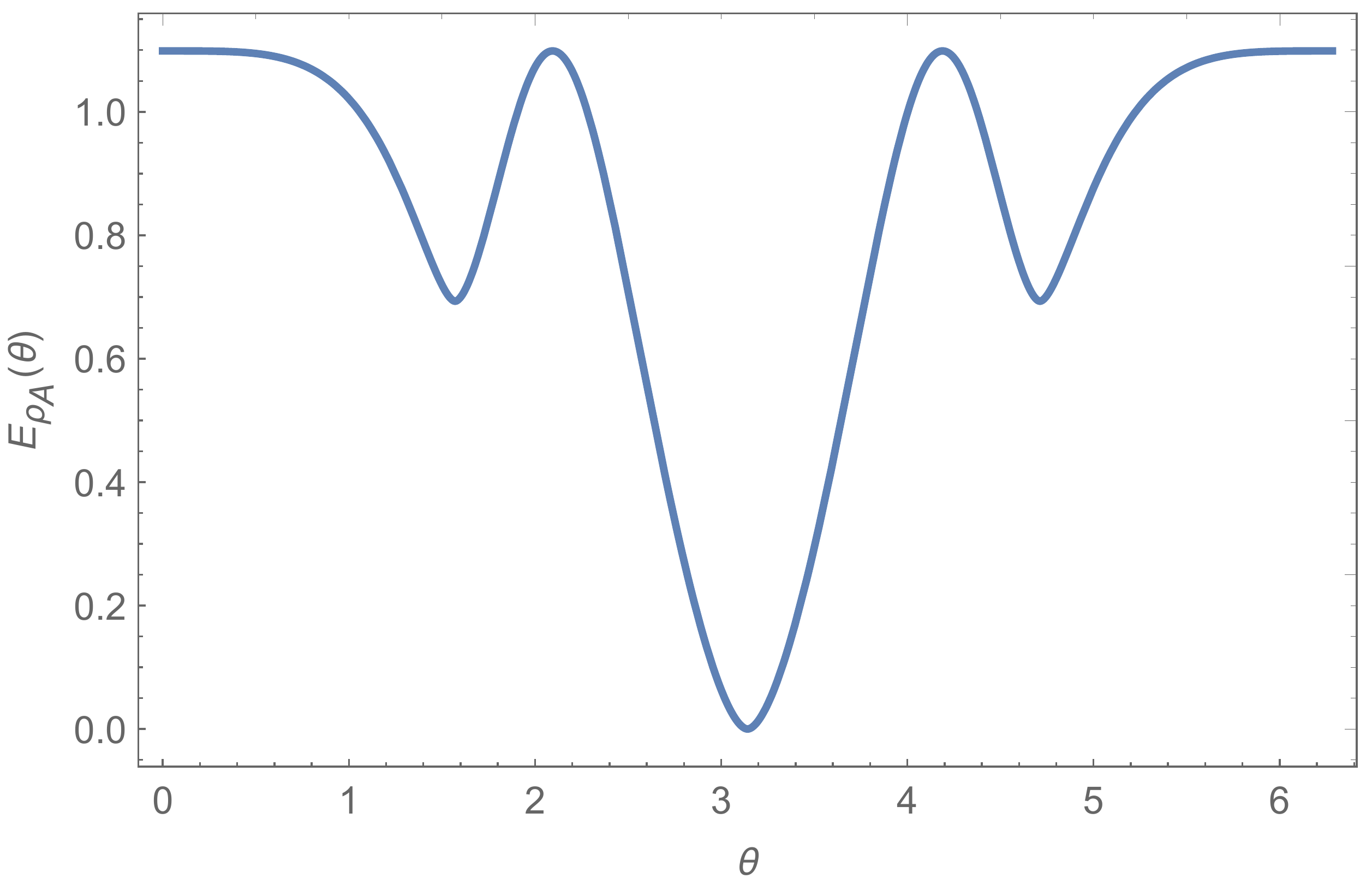}\ \hspace{0.05cm}
  \includegraphics[scale=0.26]{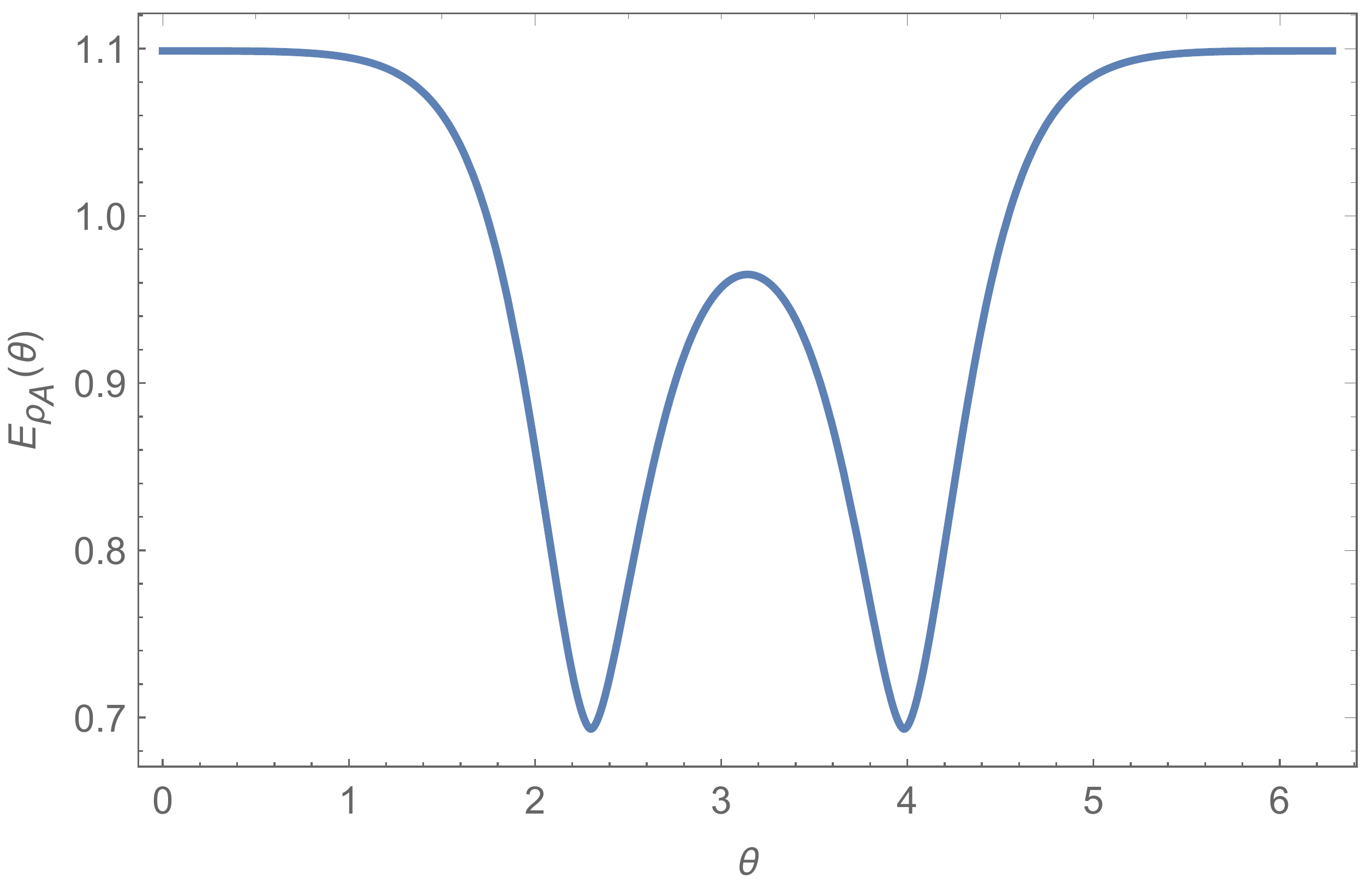}\ \hspace{0.05cm}
  \caption{\label{fig1} The sketch of the spin boundary state described by Eq.(\ref{eq:eps1}). The spins on dangling edges are fixed as $J_1=J_2=1$. The entanglement entropy as a function of $\theta$ with $j_{1}= j_{2}=\frac{1}{2}$ (upper right); $j_{1}=j_{2}=1$ (lower left); and $j_{1}=2, j_{2}=1$(lower right).}}
\end{figure}
For numerical simulation we consider a specific example as shown
in Fig.\ref{fig1}. The corresponding boundary spin state is
\begin{eqnarray}\label{eq:eps1}
\left |   \Psi _{J_{1};J_{2}}\right \rangle &=& \sum_{M_{1}M_{2}} N
C_{M_{1}\ m_{1} \ m_{2}}^{J_{1}\ j_{1} \ j_{2}} D^{m_{1}n_{1}}
D^{m_{2}n_{2}}U_{n_{1}}^{k_{1}*}(g(\theta ))U_{n_{2}}^{k_{2}*}(g(0
))C_{M_{2}\ k_{1} \ k_{2}}^{J_{2}\ j_{1} \ j_{2}}  \left | M_{1}
M_{2} \right \rangle,
\end{eqnarray}\
where $N$ is the normalization coefficient, $C_{M\ m_{1}
\ m_{1}}^{J\ j_{1} \ j_{2}}=  \left \langle j_{1}\ m_{1}; j_{2} \
m_{2}| J \ M \right \rangle$ is the standard Clebsch-Gordan
coefficient, and $D^{m n}=(-1)^{j-m} \delta ^{m ,-n}$ is the
virtual two-valent intertwiner denoting the direction of the
holonomy. $U_{n_i}^{k_i}$ is the matrix representation of the
holonomy along the edge with $j_i$. In particular, we specify the
group elements for each holonomy as $U_{n_i}^{k_i}(g(\theta )) =
e^{-ik_i\theta }\delta_{n_i}^{k_i}$, where $\theta$ is the group
parameter. For simplicity, we also ignore the holonomy along
dangling edges, where they are uniformly taken as the unit element
of $SU(2)$.

We now consider the entanglement entropy for this
bipartite system. We choose $A=\{J_1\}$ and $B=\{J_2\}$, so that the reduced density matrix is given by  $\rho _{A} = Tr
_{B}( \left | \Psi _{J_{1};J_{2}} \right \rangle \left \langle
\Psi _{J_{1};J_{2}} \right |)$. As a result, the entanglement
entropy can be evaluated as
\begin{equation}
E_{\rho _A}(\theta )= -Tr(\frac{\rho _A \ln \rho _A}{\left \langle
\Psi _{J_{1};J_{2}} |\Psi _{J_{1};J_{2}}  \right \rangle}).
\end{equation}
The numerical results for various spins are shown in
Fig.\ref{fig1}. Firstly, we note that the entanglement
entropy is not independent of the group elements any more; it is a function of the parameter $\theta$. Secondly, we
find that the entropy satisfies the bounds
$|\ln(2j_1+1)-\ln(2j_2+1)|\leq E_{\rho _A}(\theta ) \leq
\ln(2j_1+1)+\ln(2j_2+1)$. In fact, in this special case, since
there is only one dangling edge at each vertex, a stronger upper
bound holds $E_{\rho _A}(\theta ) \leq \min\{\ln(2J_1+1),
\ln(2J_2+1)\}$. It is also interesting to note that the
entanglement entropy vanishes for $\theta=\pi$ (lower left
plot of Fig.\ref{fig1}), which means that it is simply a direct
product state.

\begin{figure} [h]
  \center{
 \includegraphics[scale=0.4]{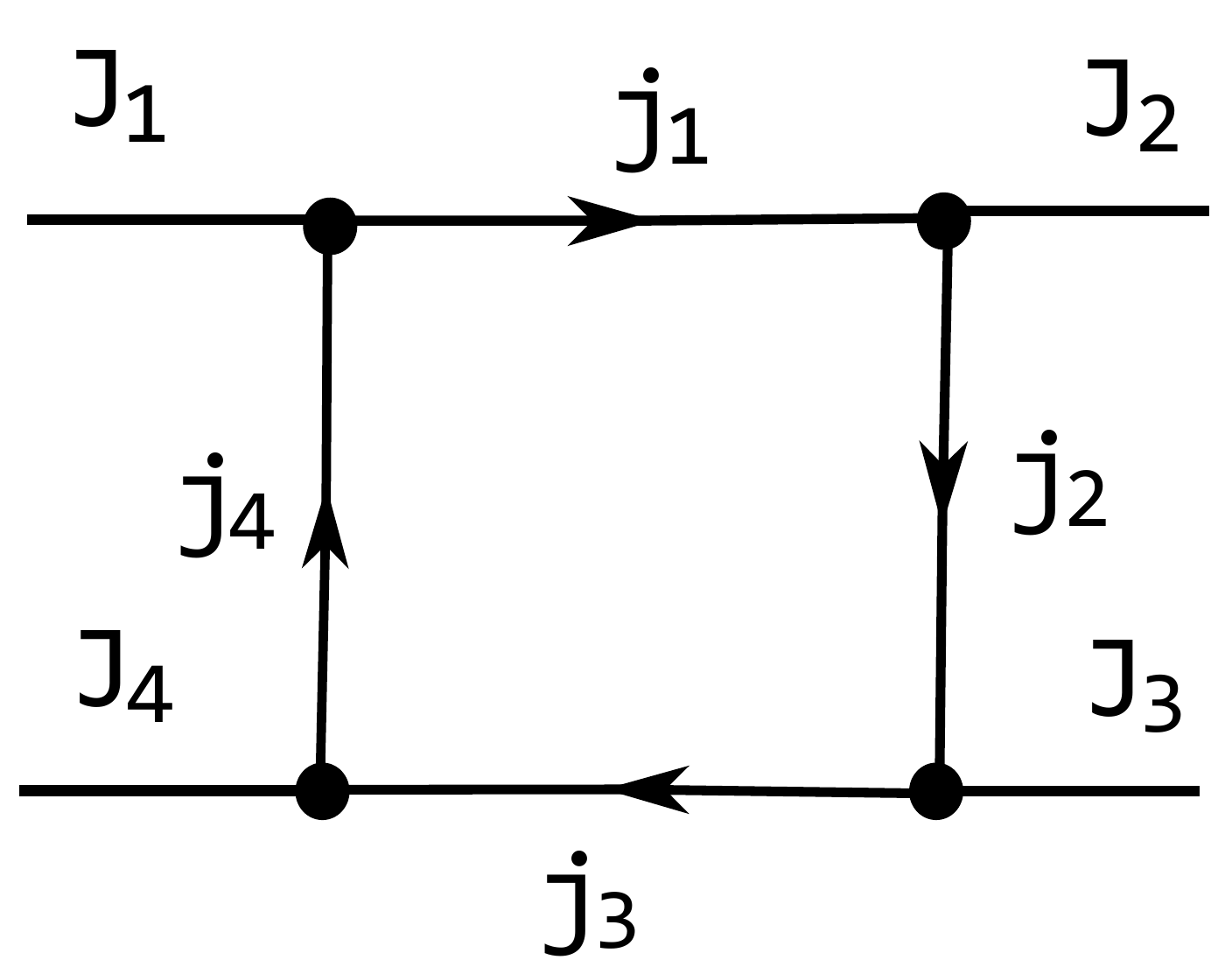}\ \hspace{0.1cm}
 \includegraphics[scale=0.24]{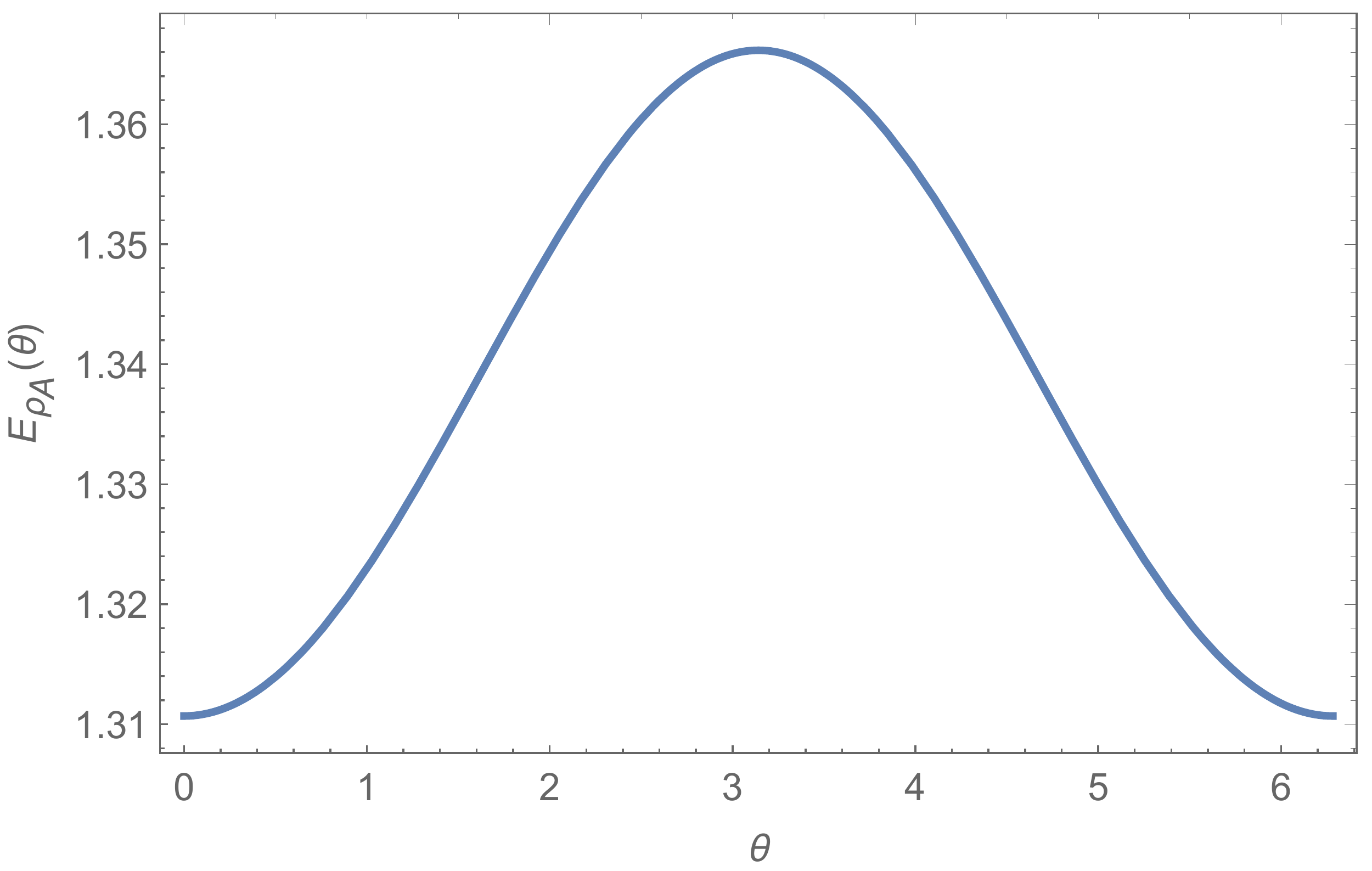}\ \hspace{0.05cm}
  \includegraphics[scale=0.24]{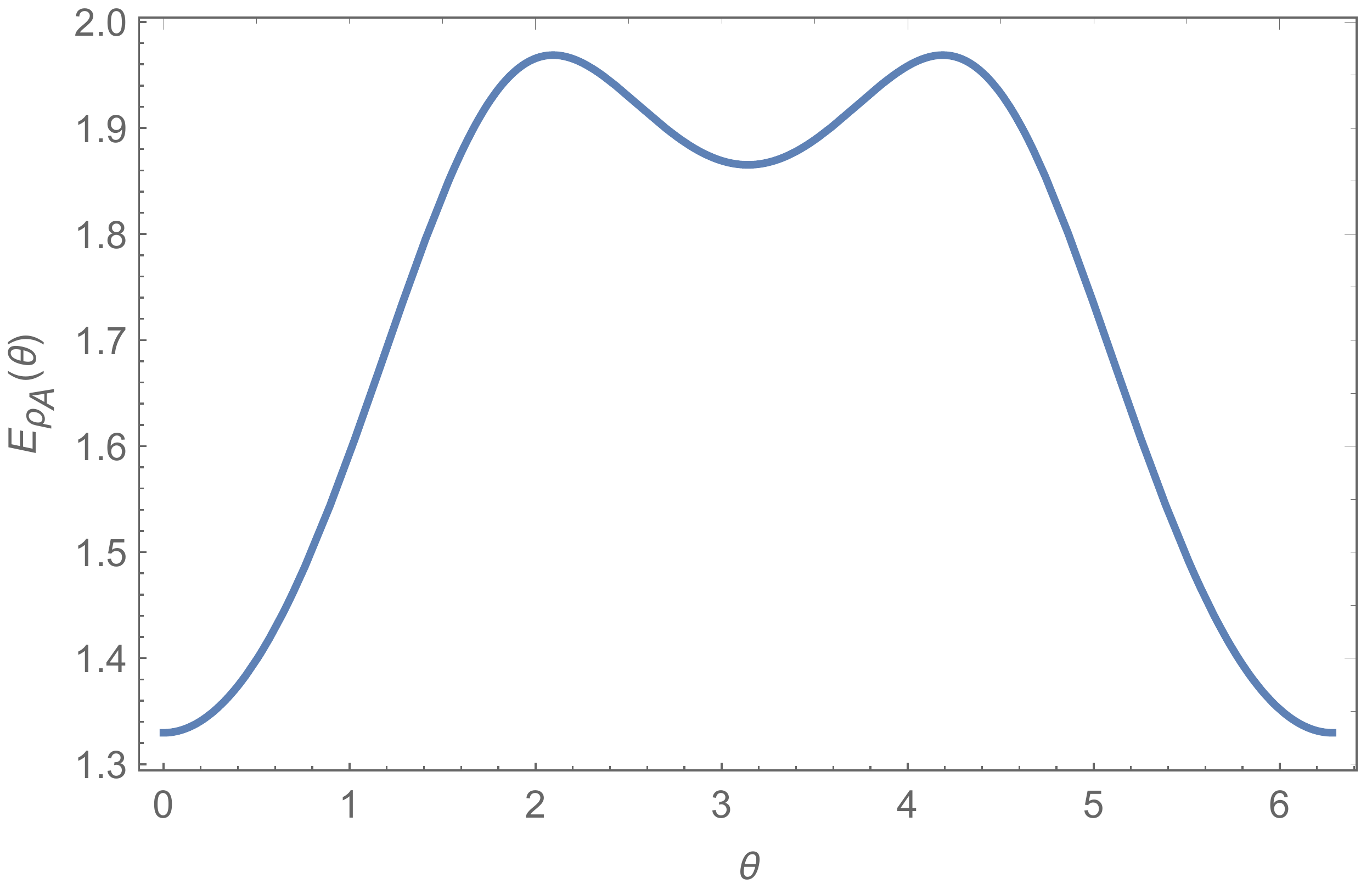}\ \hspace{0.05cm}
  \includegraphics[scale=0.24]{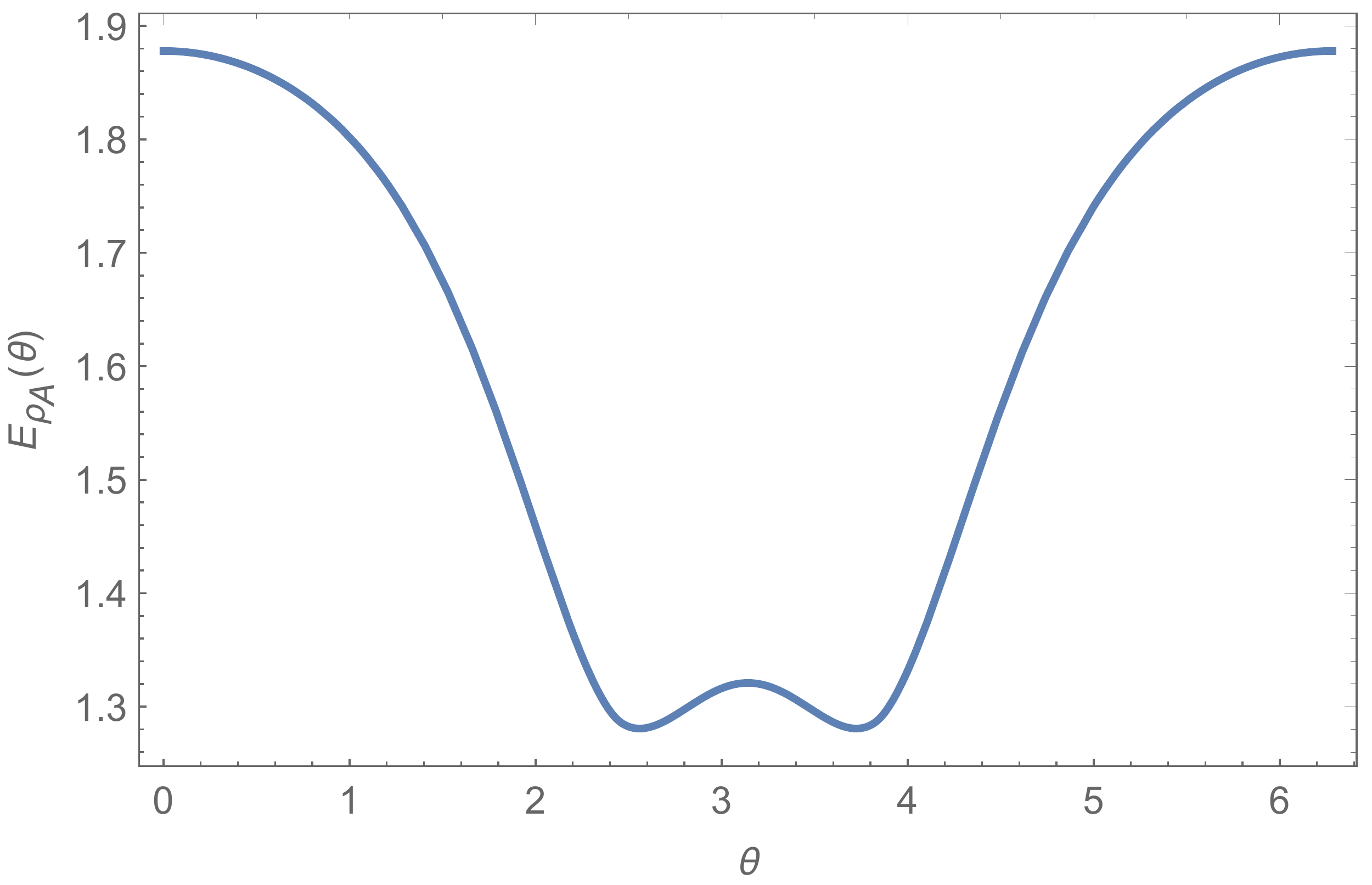}\ \hspace{0.05cm}
\caption{ The sketch of the boundary state described
by Eq.(\ref{eq:eps3}).  The spins on dangling edges have been fixed
as $J_i=1 (i=1,...4)$. The entanglement entropy as a function of
$\theta$ with $j_i=\frac{1}{2} (i=1,...4)$ (upper right); $j_i=1
(i=1,...4)$ (lower left); and $j_{1}=j_{3}=2,j_{2}=j_{4}=1$ (lower
right).}
\label{fig5}
  }
\end{figure}

Next, we consider the case that two vertices are linked by more
than one path, which means that some paths may connect them
indirectly by passing through other vertices. This is, of course,
a common case for general spin networks. As an example, we
consider the spin network shown in Fig.\ref{fig5}. The
corresponding boundary state is given as
\begin{eqnarray}\label{eq:eps3}
\left |   \Psi _{J_{1} J_{4};J _{2} J _{3} }\right \rangle &=& \sum_{M_{l}} NC_{M_{1}\ m_{1} \ k_{4}}^{J_{1}\ j_{1} \ j_{4}} D^{m_{1}n_{1}}  U_{n_{1}}^{k_{1}*}(g(\theta ))  C_{M_{2}\ m_{2} \ k_{1}}^{J_{1}\ j_{2} \ j_{1}} D^{m_{2}n_{2}}  U_{n_{2}}^{k_{2}*}(g(0 ))  \nonumber \\
&& \times  C_{M_{3}\ m_{3} \ k_{2}}^{J_{3}\ j_{3} \ j_{2}}
D^{m_{3}n_{3}}  U_{n_{3}}^{k_{3}*}(g(0 ))   C_{M_{4}\ m_{4} \
k_{3}}^{J_{4}\ j_{4} \ j_{3}} D^{m_{4}n_{4}}
U_{n_{4}}^{k_{4}*}(g(0))
  \left | M_{1} M_{2}  M_{3} M_{4}\right \rangle,
\end{eqnarray}
where, for a bipartite system, we have chosen $A=\{J_1,J_4\}$ and
$B=\{J_2,J_3\}$ and $l =1, \cdots 4$.The reduced
density matrix for the bipartite entanglement entropy is given as
$\rho _{A} = Tr _{B}( \left | \Psi _{J _{1} J_{4};J _{2} J _{3} }
\right \rangle \left \langle \Psi _{J _{1}J_{4}; J _{2} J _{3}}
\right |)$.Numerical results for a few specific spins are
shown in Fig.\ref{fig5}. We note that the entanglement
entropy is generally a function of the parameter $\theta$. In
particular, when two parts are linked by two edges with spins
$j_1$ and $j_3$, respectively, we find that the entropy is bounded as $|\ln(2j_1+1)-\ln(2j_3+1)|\leq E_{\rho _A}(\theta )
\leq \ln(2j_1+1)+\ln(2j_3+1)$.

\section{Entanglement in the presence of intertwiner entanglement}\label{sec-epe}

  \begin{figure} [t]
  \center{
  \includegraphics[scale=0.7]{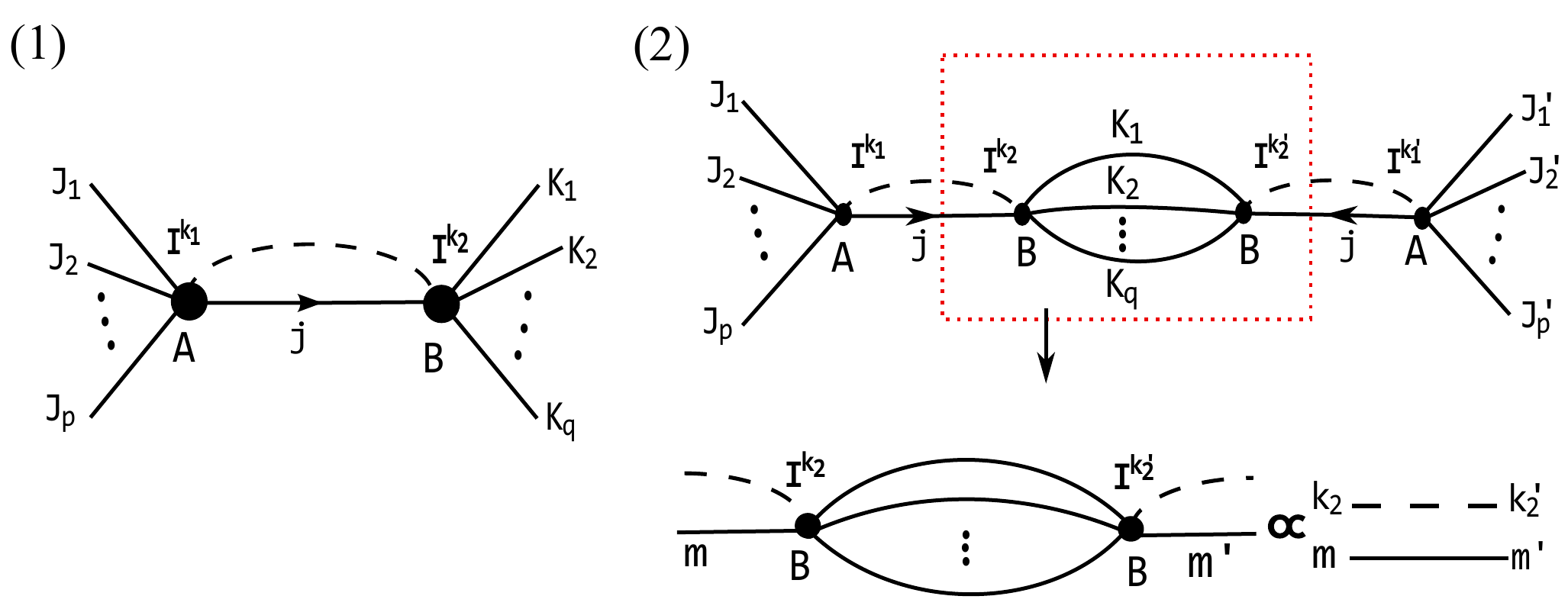}\ \hspace{0.1cm}
  \caption{ (1) Two neighboring
vertices $A$ and $B$ are linked by a single edge carrying a spin
$j$. The dashed line denotes the entanglement of two
interwiners. (2) The sketch of the orthogonal relation
described by Eq.(\ref{eq:eps4}).} \label{fig14}
  }
  \end{figure}

In this section, we take the intertwiner entanglement into account.
In ref. \cite{Livine:2017fgq}, it was shown that when two neighboring
vertices $A$ and $B$ are linked by a single edge carrying a spin
$j$, as shown in Fig.\ref{fig14}(1), then the total
entanglement entropy of the boundary states can be separated into
two parts, one from intertwiner entanglement at vertices and the
other from spin entanglement, which is nothing but $\ln(2j+1)$, the
maximal entropy allowed by the spin on the edge and independent of
the group elements of the holonomy.We point
out that the following relation plays a crucial role in
the separation of spin entanglement and intertwiner entanglement; it is the orthogonal relation between two intertwiners
\begin{eqnarray}\label{eq:eps4}
\sum_{N_{1}\cdots N_{q}}\left \langle I^{k_2} | N_{1}\cdots N_{q}m
\right \rangle\left \langle N_{1}\cdots N_{q}m' | I^{k'_2} \right
\rangle &=& \frac{1}{2j+1}\delta _{k_2k'_2}\delta _{mm'},
\end{eqnarray}
where $\left |I^{k_2} \right \rangle$ represents the $k_2$-th
component of the intertwiner state, and $N_i$ ($i=1, \cdots, q$) is
the magnetic quantum number of the spin on the $i$-th dangling
edge, while $m$ is the magnetic quantum number of the spin $j$ on
the single edge linking two vertices.

This orthogonal relation can be represented as a diagram, as shown
in Fig.\ref{fig14}(2). Obviously, this identity is applied during
the evaluation of the reduced density matrix such that the final
result can be written as a product of the spin contribution and
the intertwiner contribution, as shown in Eq.(24) in
\cite{Livine:2017fgq}.

  \begin{figure} [t]
  \center{
  \includegraphics[scale=0.7]{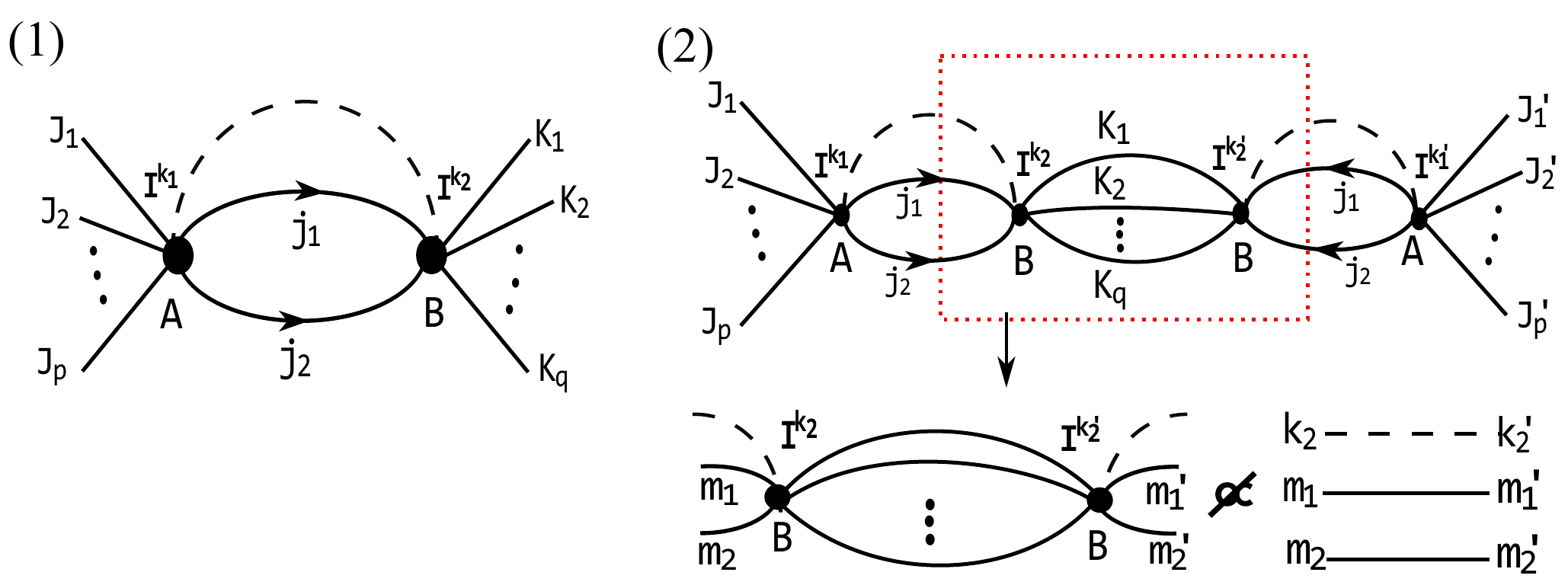}\ \hspace{0.1cm}
  \caption{ (1) Two neighboring
vertices $A$ and $B$ are linked by two edges carrying spin $j_{1}$
and $j_{2}$, respectively. (2) The extension of the
orthogonal relation does not hold.
  }
  \label{fig15}}
  \end{figure}

However, when two vertices are linked by two or more edges, we
find that this situation does not hold any more. In general, the
bipartite entanglement entropy can not be separated into a spin part
and an intertwiner part. For explicitness, we consider two vertices
$A$ and $B$ linked by two edges carrying spin $j_{1}$ and $j_{2}$,
respectively, as shown in Fig.\ref{fig15}(1). For
the evaluation of the reduced density matrix, we need to simplify the
contractions of tensors. Unfortunately, we find that the following
identity, need to separate the intertwiner entanglement
from spin entanglement, does not hold,
\begin{small}
\begin{eqnarray}\label{eq:eps8}
\sum_{N_{1}\cdots N_{q}}\left \langle I^{k_2} |N_{1}
\cdots N_{q} m_{1}m_{2}\right \rangle \left \langle N_{1} \cdots
N_{q} m'_{1}m'_{2}| I^{k'_2} \right \rangle &\neq&
\frac{1}{(2j_{1}+1)(2j_{2}+1)} \delta _{k_2k'_2}\delta
_{m_{1}m'_{1}}\delta _{m_{2}m'_{2}}.
\end{eqnarray}
\end{small}
This is diagrammatically sketched in Fig.\ref{fig15}(2). We
provide the proof for this statement in the Appendix. Similarly,
one can show that such relations are also absent when two
vertices are linked by more than two edges indirectly. Therefore,
for a general spin network with dangling edges, it is not possible
to separate the total entropy into the  contributions from spins on
the edges and from intertwiners at vertices.

\begin{figure} [t]
  \center{
  \includegraphics[scale=0.7]{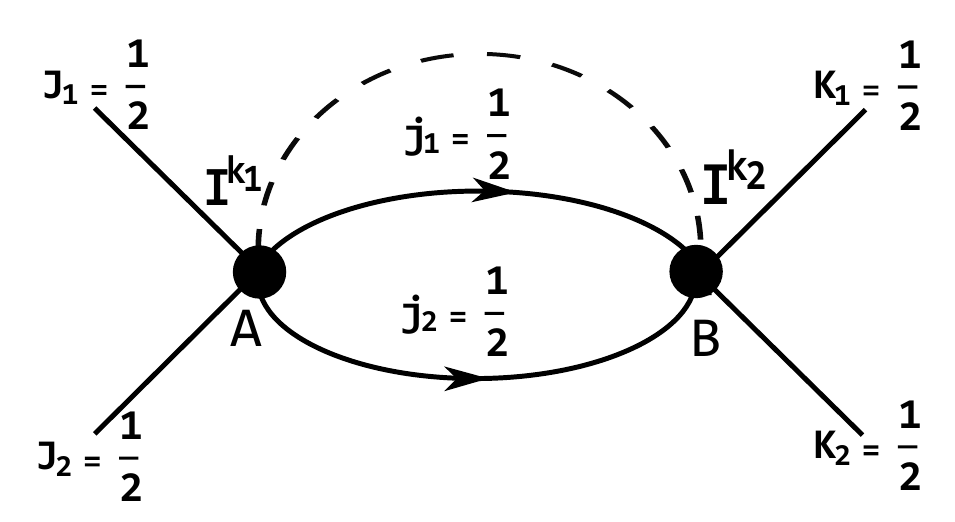}\ \hspace{0.1cm}
  \caption{ The sketch of the boundary spin state with intertwiner entanglement described by Eq.(\ref{eq:eps22}). }
  \label{fig16}}
  \end{figure}

The above orthogonal relation is not a necessary condition
for separating the intertwiner indices and spin indices. However, we
remark that, in a general case, they can not be separated if two of vertices are linked by more than one path. To support this statement, we evaluate the total
entanglement entropy and the intertwiner entanglement entropy
numerically for a few specific spin networks. An example is
shown in Fig.\ref{fig16}, and the boundary spin state reads
\begin{eqnarray}\label{eq:eps22}
\left | \Psi \right \rangle=\sum_{M_{i} N_{i} } \frac{\varphi _{k_{1}k_{2}} }{\sqrt{(2 k_{1} +1)(2 k_{2} +1)} }  C_{n_{1}\ M_{1} \ M_{2}}^{k_{1} \ J_{1} \ J_{2}} C_{n'_{1}\ m_{1} \ m_{2}}^{k_{1} \ j_{1} \ j_{2}} D^{n_{1}n'_{1}}   C_{n'_{2}\ m'_{1} \ m'_{2}}^{k_{2} \ j_{1} \ j_{2}} C_{n_{2}\ N_{1} \ N_{2}}^{k_{2} \ K_{1} \ K_{2}}
D^{n_{2}n'_{2}}  D^{m_{1}m'_{1}} D^{m_{2}m'_{2}} \left | M_{1}
M_{2} N_{1} N_{2}   \right \rangle,
\end{eqnarray}
where $i = 1, 2$ and $k_{1},k_{2}$ are possible spins on
virtual edges inside intertwiner $A$ and $B$, respectively.

If we take all spins on the dangling edges to be $\frac{1}{2}$,
then the spins on virtual edges inside an intertwiner can be $0$
and $1$. With this assumption, the boundary state takes the following
general form,
\begin{eqnarray}\label{eq:eps23}
\left | \Psi \right \rangle &=& \sum_{M_{i} N_{i} } ( \varphi _{00} C_{0 \ M_{1} \
M_{2}}^{0 \ \frac{1}{2} \ \frac{1}{2}}C_{0 \ N_{1} \ N_{2}}^{0 \
\frac{1}{2} \ \frac{1}{2}} + \frac{1}{3} \varphi _{11}  C_{n_{1} \
M_{1} \ M_{2}}^{1 \ \frac{1}{2} \ \frac{1}{2}} C_{n_{2} \ N_{1} \
N_{2}}^{1 \ \frac{1}{2} \ \frac{1}{2}} D^{n_{1}n_{2}} ) \left |
M_{1} M_{2} N_{1} N_{2}   \right \rangle,
\end{eqnarray}
 where $\varphi _{00}$ and $\varphi _{11}$ are two components in the
 intertwiner space. It should be noted that the other two
 components $\varphi _{01}$ and $\varphi _{10}$ do not appear in the
 above equation simply because the contraction of the corresponding $CG$
 coefficients in these terms vanishes.

The reduced density matrix for bipartition is given as $\rho
_{M_{1}M_{2}} =Tr_{N_{1}N_{2}}(\left | \Psi \right \rangle \left
\langle \Psi \right |)$. It is straightforward to obtain the
entanglement entropy, which is
\begin{eqnarray}\label{eq:eps24}
E &=& -Tr(\frac{\rho_{M_{1}M_{2}} \ln \rho_{ M_{1}M_{2}}}{\left \langle \Psi  | \Psi  \right \rangle}) \nonumber \\
              &=& -\frac{ 3 \left |\varphi _{00}  \right |^{2} }{3\left |\varphi _{00}  \right |^{2} + \left |\varphi _{11}  \right |^{2} } \ln (\left|\varphi _{00}  \right |^{2}) - \frac{  \left |\varphi _{11}  \right |^{2} }{3\left |\varphi _{00}  \right |^{2} + \left |\varphi _{11}  \right |^{2} } \ln (\left|\varphi _{11}  \right |^{2}) \nonumber \\
               &&+ \frac{  \left |\varphi _{11}  \right |^{2} - 3 \left |\varphi _{00}  \right |^{2} }{3\left |\varphi _{00}  \right |^{2} + \left |\varphi _{11}  \right |^{2} } \ln 3  + \ln( 3\left |\varphi _{00}  \right |^{2} + \left |\varphi _{11}  \right |^{2}
               ).
         \end{eqnarray}

On the other hand, the intertwiner entanglement entropy is
determined by the matrix $\varphi _{k_{1}k_{2}}$,
\begin{equation}  \label{eq:eps25}
          \varphi _{k_{1}k_{2}} = \left(
             \begin{array}{cc}
          \varphi _{00} & \varphi _{01}\\
          \varphi _{10} & \varphi _{11} \\
          \end{array}
       \right).
\end{equation}
The reduced density matrix is  $\rho_{k_{1}}=\frac{\varphi _{k_{1}
k_{2}} \varphi _{k_{1} k_{2}}^{\dagger }}{Tr( \varphi _{k_{1}
k_{2}}^{\dagger} \varphi _{k_{1} k_{2}}  )} $. The entanglement
entropy between intertwiners is
\begin{eqnarray}\label{eq:eps26}
E_I = - Tr( \rho_{k_{1}} \ln \rho_{k_{1}} ) = -(a_{+} \ln a_{+} +
a_{-} \ln a_{-}),
\end{eqnarray}
where $a_{\pm } = (\frac{1}{2} \pm  \frac{1}{2} \sqrt{1-  \frac{4
\left | \varphi _{00}\varphi _{11} - \varphi _{01}\varphi _{10}
\right |^{2}}{  \left | \varphi _{00} \right |^{2} + \left |
\varphi _{01} \right |^{2} + \left | \varphi _{10} \right |^{2} +
\left | \varphi _{11} \right |^{2}   } }) $. In Tab.\ref{tab}, we
evaluate the entanglement entropy $E$ of the boundary spin state
and the entanglement entropy $E_I$ of intertwiners for a few
specific values of intertwiner parameters. It manifestly indicates
that the total entanglement entropy measured in boundary states
can not be written as the sum of the spin contribution and the
intertwiner contribution. For instance, in the fifth
column of the table, the entanglement entropy between intertwiners
is even larger than the entanglement entropy for the boundary
state. In the last column, the total entanglement entropy of the
boundary state is zero, but the entanglement of intertwiners is
not. In the next-to-last column, ``meaningless'' means that
the boundary state $| \Psi \rangle$ vanishes. Finally, we remark
that the total entanglement entropy is not larger than
$\ln4$ in
all cases considered, simply because all dangling edges carry spin
$1/2$. In general, the bounds we found in the previous section do
not hold any more when the intertwiner entanglement is involved.
\begin{table} [h]
       \center{

         \resizebox{\textwidth}{10mm}{
       \begin{tabular}{|c|c|c|c|c|c|c|c|p{3cm}<{\centering}| }
         \hline
          $\varphi _{k_{1}k_{2}}$ &  $\left( \begin{array}{cc}
  1 & 0\\
   0 & 0 \\
 \end{array} \right )$ & $\left( \begin{array}{cc}
  1 & 0\\
   0 & 1 \\
 \end{array} \right )$ &
  $\left( \begin{array}{cc}
  1 & 0\\
   0 & 3 \\
 \end{array} \right )$ &
 $\left( \begin{array}{cc}
  3 & 0\\
   0 & 1 \\
 \end{array} \right )$ &

 $\left( \begin{array}{cc}
  1 & 1\\
   1 & 1 \\
 \end{array} \right )$  &
 $\left( \begin{array}{cc}
  1 & \sqrt{3}\\
  \sqrt{3} & 3 \\
 \end{array} \right )$&
 $\left( \begin{array}{cc}
  0 & 1\\
   1 & 0 \\
 \end{array} \right )$  &
 $\left( \begin{array}{cc}
  1 & 1\\
   1 & 0 \\
 \end{array} \right )$ \\ \hline
       \multirow{2}{*}{ $E$}& \multirow{2}{*} {0} & \multirow{2}{*} {$\ln 4 - \frac{1}{2} \ln3 $}& \multirow{2}{*}{$\ln 4 $} & \multirow{2}{*}{ $\ln 28 - \frac{20}{7} \ln3 $} &\multirow{2}{*}{ $\ln 4 - \frac{1}{2} \ln3$ } & \multirow{2}{*}{$\ln 4  $} & \multirow{2}{*}{meaningless } & \multirow{2}{*} {0} \\

        &  & &  &   &  &   &  & \\ \hline

     \multirow{2}{*} {$E_I$}& \multirow{2}{*} {$0$}  & \multirow{2}{*} {$\ln2$} &\multirow{2}{*} {$ \ln10 - \frac{9}{5} \ln 3$}  & \multirow{2}{*} {$ \ln10 - \frac{9}{5} \ln 3$} & \multirow{2}{*} {0} & \multirow{2}{*} {0} & \multirow{2}{*} {$\ln2$} &  $-\frac{3 + \sqrt{5}}{6} \ln(\frac{3 + \sqrt{5}}{6}) -  \frac{3 - \sqrt{5}}{6} \ln(\frac{3 - \sqrt{5}}{6})$ \\ \hline
      \end{tabular}}

\caption{The entanglement entropy E of boundary spin state
and the entanglement entropy $E_I$ of intertwiners for
various intertwiner matrices.}\label{tab} }
 \end{table}

\section{Conclusions and outlook} \label{sec-cao}
In this paper, we have investigated the bipartite entanglement for
the boundary spin states in spin networks with dangling edges. In
particular, we have constructed a simple type of spin network in
which two complementary parts are linked by two paths, either in a
direct or indirect manner. The numerical evaluation of
entanglement entropy leads to the following two main results.
Firstly, in the absence of the intertwiner entanglement, the
entanglement entropy for the boundary state depends on the group
elements of the holonomy, which can not be simply determined by
the spins $j_1$ and $j_2$ on the edges connecting the complementary
parts. Nevertheless, we have proposed a bound for the entanglement
entropy, which is $|\ln(2j_1+1)-\ln(2j_2+1)|\leq E \leq
\ln(2j_1+1)+\ln(2j_2+1)$. It would be very important to prove or test
this bound in a general case. Secondly, when the intertwiner
entanglement is taken into account, the total
entanglement can not be written, in general, as the sum of intertwiner
entanglement and spin entanglement, but as a mixture of these two
contributions.

Although we have only considered the simple case with two paths
connecting two vertices, we believe that the above statements could be
applicable to more complicated cases in which two vertices are
linked by more than two edges directly, or by indirect paths.

Finally, based on our current work it is quite intriguing to
further explore the relationship between quantum entanglement and
quantum geometry, described by spin network states in loop
quantum gravity. Our investigation is in progress and will
be published in the near future\cite{LWX}.

\section{Appendix}\label{sec-a}

  \begin{figure} [h]
  \center{
  \includegraphics[scale=0.7]{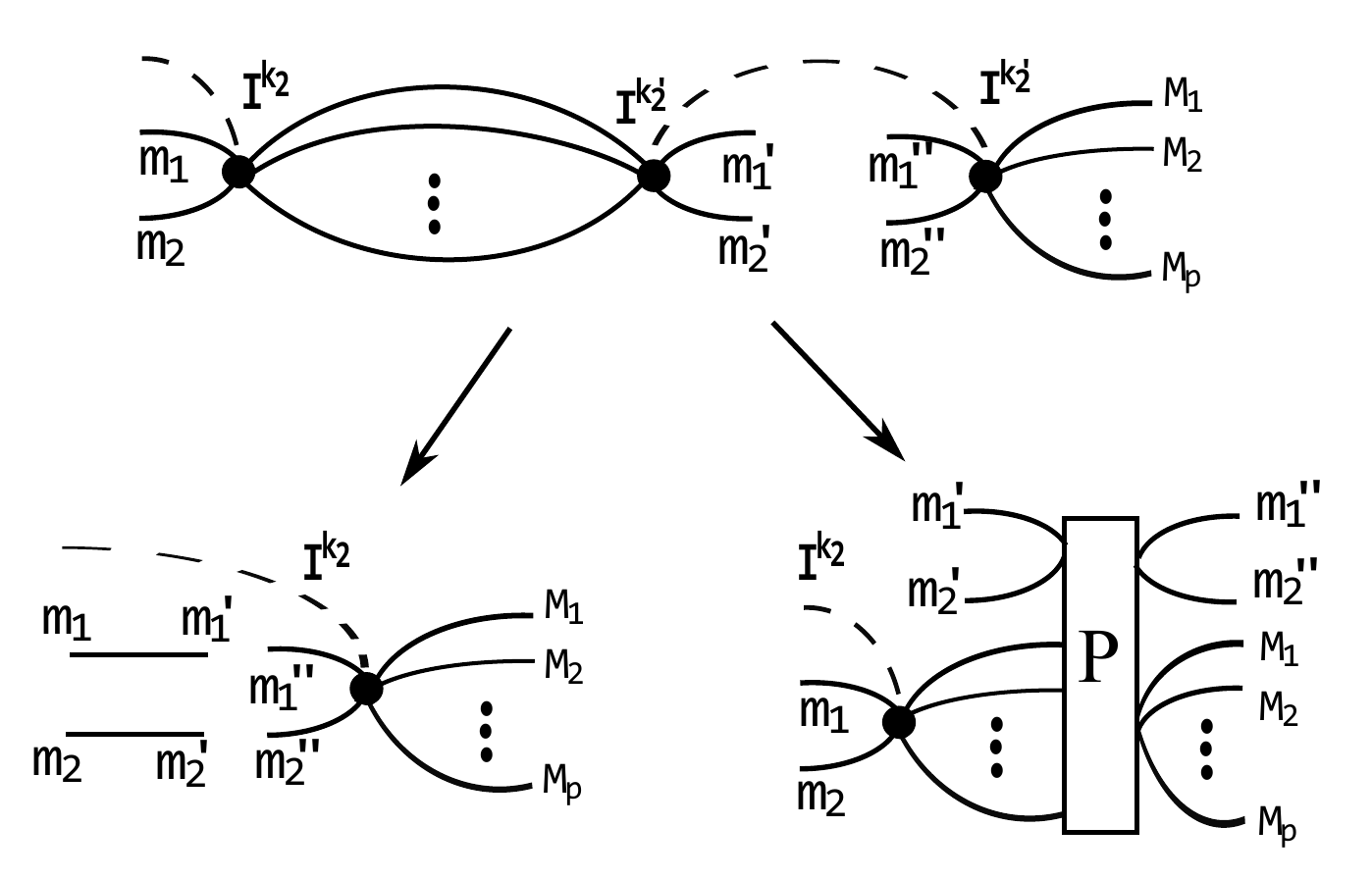}\ \hspace{0.1cm}
  \caption{\label{fig17} The diagrammatic sketch for the processes in Eqs.(\ref{eq:eps9}), (\ref{eq:eps10}), (\ref{eq:eps11}) .
  }}
  \end{figure}

In this Appendix, we demonstrate the absence of the orthogonal relation for
intertwiners when two vertices are linked by two edges, namely the
inequality in Eq.(\ref{eq:eps8}), by applying the proof by
contradiction.  Assume that Eq.(\ref{eq:eps8}) is true. Let us
consider the following contraction, which appears in the evaluation
of the reduced density matrix
\begin{eqnarray}\label{eq:eps9}
D &=&
\sum_{N_{1}\cdots N_{q}} \left \langle I^{k_2} |
N_{1}\cdots N_{q} m_{1}m_{2} \right \rangle \sum_{k'_2} \left \langle
 N_{1}\cdots N_{q} m'_{1}m'_{2} | I^{k'_2}  \right \rangle \left
\langle I^{k'_2} | M_{1}\cdots M_{p} m''_{1}m''_{2} \right \rangle.
\end{eqnarray}
From Eq.(\ref{eq:eps8}), one can write Eq.(\ref{eq:eps9}) as,
\begin{eqnarray}\label{eq:eps10}
D &=&
\frac{1}{(2j_{1}+1)(2j_{2}+1)} \delta _{m_{1}m'_{1}} \delta
_{m_{2}m'_{2}} \left \langle I^{k_2} |  M_{1}\cdots M_{p} m''_{1}m''_{2}
\right \rangle.
\end{eqnarray}

For convenience, we define the operator $\widehat{P} =\sum \left |
I^{k'_2} \right \rangle \left \langle I^{k'_2} \right |$, so that Eq.(\ref{eq:eps9}) can be rewritten as
\begin{eqnarray}\label{eq:eps11}
D &=&
\sum_{N_{1}\cdots N_{q}} \left \langle I^{k_2} |
N_{1}\cdots N_{q}  m_{1}m_{2} \right \rangle  \left \langle
 N_{1}\cdots N_{q}m'_{1}m'_{2} | \widehat{P} | M_{1}\cdots M_{p}  m''_{1}m''_{2}\right \rangle.
\end{eqnarray}

\begin{figure} [t]
  \center{
  \includegraphics[scale=0.7]{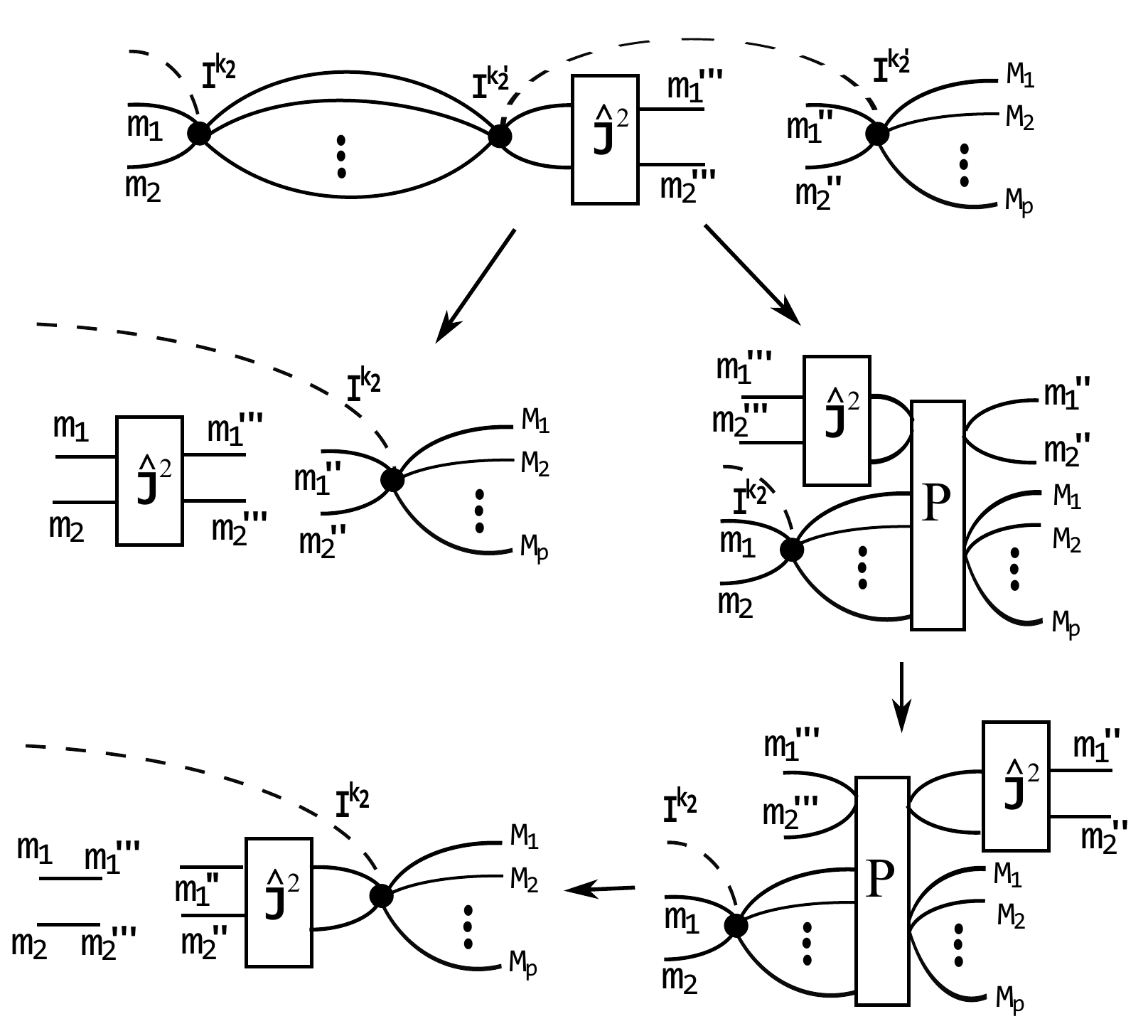}\ \hspace{0.1cm}

  \caption{\label{fig18} The diagrammatic sketch for the processes in Eqs. (\ref{eq:eps13}), (\ref{eq:eps14}), (\ref{eq:eps15}), (\ref{eq:eps16}), (\ref{eq:eps17}).
  }}
  \end{figure}

A diagrammatic sketch of Eqs.(\ref{eq:eps9}), (\ref{eq:eps10}) and
(\ref{eq:eps11}) is shown in Fig.\ref{fig17}. We
introduce the operator $ \widehat{J}^{2} = (\widehat{J}_{1} +
\widehat{J}_{2} )^{2}$. The action of operators $\widehat{J}_{1}$
and $\widehat{J}_{2}$ is defined as
\begin{eqnarray}\label{eq:eps12}
 \left \langle m_{1}m_{2}|\widehat{J}_{1} | m'_{1}m'_{2} \right \rangle &=& \left \langle m_{1}|\widehat{J}_{1} | m'_{1} \right \rangle \delta _{m_{2}m'_{2}} \nonumber \\
 \left \langle m_{1}m_{2}|\widehat{J}_{2} | m'_{1}m'_{2} \right \rangle &=& \delta _{m_{1}m'_{1}} \left \langle m_{2}|\widehat{J}_{2} | m'_{2} \right
 \rangle,
\end{eqnarray}
where $\widehat{J}_{i}^2 \left | j_i m \right \rangle = j_i(j_i +
1)\left | j_i m \right \rangle $ ($i = 1,2$). Next, we consider the
following action of this operator on $D$, which is denoted as $F$
and shown in Fig.\ref{fig18}.

\begin{eqnarray}\label{eq:eps13}
F &=& \sum_{N_{1}\cdots N_{q}} \left \langle I^{k_2} |
N_{1}\cdots N_{q}  m_{1}m_{2} \right \rangle \sum_{m'_1 m'_2} \left
\langle m'''_{1} m'''_{2} |\widehat{J}^{2}| m'_{1}m'_{2} \right \rangle  \left \langle
 N_{1}\cdots N_{q}m'_{1}m'_{2} | \widehat{P} |  M_{1}\cdots M_{p}m''_{1}m''_{2} \right \rangle.
\end{eqnarray}

On the one hand, by virtue of Eq.(\ref{eq:eps10}), Eq.(\ref{eq:eps13}) can be simplified as
\begin{eqnarray}\label{eq:eps14}
F &=&
\frac{1}{(2j_{1}+1)(2j_{2}+1)} \left \langle m'''_{1} m'''_{2}
|\widehat{J}^{2}| m_{1}m_{2} \right
\rangle  \left \langle I^{k_2}|M_{1}\cdots M_{p} m''_{1}m''_{2}  \right
\rangle.
\end{eqnarray}
On the other hand, from Eq.(\ref{eq:eps11}), we may rewrite Eq.(\ref{eq:eps13}) as
\begin{eqnarray}\label{eq:eps15}
F&=& \sum_{N_{1}\cdots N_{q}} \left \langle I^{k_2} | m_{1}m_{2}
N_{1}\cdots N_{q}  \right \rangle
\sum_{m'_{1}m'_{2}} \left \langle m'''_{1}m'''_{2}
|\widehat{J}^{2} | m'_{1}m'_{2} \right \rangle  \left \langle
 N_{1}\cdots N_{q}m'_{1}m'_{2} |\widehat{P} |
M_{1}\cdots M_{p}m''_{1}m''_{2} \right \rangle.
\end{eqnarray}

Next, we prove that the operators $\widehat{J}^{2}$ and
$\widehat{P}$ commute with each other. For any $\left | \Psi
\right \rangle \in \boldsymbol{H }_{1} \otimes \boldsymbol{H }_{2}
\otimes \cdots \boldsymbol{H }_{p}$, we have $\widehat{P}\left |
\Psi  \right \rangle \in \textup{Inv}_{SU(2)} [ \boldsymbol{H
}_{1} \otimes \boldsymbol{H }_{2} \otimes \cdots \boldsymbol{H
}_{p} ]$. We also know that $\left [ \widehat{J}^{2},
\widehat{J}_{1} + \widehat{J}_{2}+\cdots \widehat{J}_{p} \right ]
= 0$ and $(\widehat{J}_{1} + \widehat{J}_{2}+\cdots
\widehat{J}_{p} ) \widehat{P}\left | \Psi \right \rangle = 0$. So,
we have $(\widehat{J}_{1} + \widehat{J}_{2}+\cdots \widehat{J}_{p}
) \widehat{J}^{2} \widehat{P}\left | \Psi  \right \rangle = 0$.
That means $\widehat{J}^{2}  \widehat{P}\left | \Psi  \right
\rangle \in \textup{Inv}_{SU(2)} [ \boldsymbol{H }_{1} \otimes
\boldsymbol{H }_{2} \otimes \cdots \boldsymbol{H }_{p} ] $. We conclude that $\widehat{P} \widehat{J}^{2}  \widehat{P}\left |
\Psi  \right \rangle = \widehat{J}^{2}
\widehat{P}\left | \Psi  \right \rangle $. Because $\left | \Psi
\right \rangle$ is arbitrary, we get $\widehat{P}
\widehat{J}^{2}  \widehat{P} = \widehat{J}^{2}  \widehat{P}$. If
we take its transposed-conjugate $\widehat{P} \widehat{J}^{2}
\widehat{P} = \widehat{P} \widehat{J}^{2} $, we get $
\widehat{J}^{2} \widehat{P} = \widehat{P} \widehat{J}^{2}  $, i.e.
$\left [ \widehat{P}, \widehat{J}^{2} \right ] = 0$. With this
fact, Eq.(\ref{eq:eps15}) becomes
\begin{eqnarray}\label{eq:eps16}
F &=&  \sum_{N_{1}\cdots N_{q}} \left \langle I^{k_2} |
N_{1}\cdots N_{q}  m_{1}m_{2} \right \rangle \sum_{m'_1 m'_2} \left \langle
N_{1}\cdots N_{q} m'''_{1}m'''_{2} | \widehat{P} |  M_{1}\cdots M_{p}m'_{1}m'_{2} \right \rangle  \left
\langle m'_{1} m'_{2} |\widehat{J}^{2}| m''_{1}m''_{2} \right \rangle.
\end{eqnarray}

With the help of Eq.(\ref{eq:eps10}) and
Eq.(\ref{eq:eps11}), the above equation can be further
simplified as
\begin{eqnarray}\label{eq:eps17}
F &=& \frac{1}{(2j_{1}+1)(2j_{2}+1)} \delta _{m_{1}m'''_{1}} \delta _{m_{2}m'''_{2}} \sum_{ m'_{1}m'_{2}} \left \langle I^{k_2} | m'_{1}m'_{2}
M_{1}\cdots M_{p}   \right \rangle \left \langle  m'_{1}m'_{2}
|\widehat{J} ^{2}| m''_{1}m''_{2}
\right \rangle.
\end{eqnarray}
A diagrammatic sketch of Eqs.
(\ref{eq:eps13})-(\ref{eq:eps17}) is shown in
Fig.\ref{fig18}.

If we contract both  Eq.(\ref{eq:eps14}) and Eq.(\ref{eq:eps17})
with $\left \langle M_{1}\cdots M_{p}  m''_{1}m''_{2}| I^{k_2}
\right \rangle $, we get
\begin{eqnarray}\label{eq:eps18}
&&\left \langle m'''_{1} m'''_{2}|\widehat{J}^{2}| m_{1}m_{2} \right\rangle \sum_{k_2, m''_{i}, M_{l}} \left \langle I^{k_2}|M_{1}\cdots M_{p} m''_{1}m''_{2}  \right\rangle \left \langle  M_{1}\cdots M_{p} m''_{1}m''_{2}| I^{k_2}  \right\rangle = \delta _{m_{1}m'''_{1}} \delta _{m_{2}m'''_{2}}  \nonumber \\
&& \times \sum_{k_2, m'_{i},m''_{i},M_{l} } \left \langle I^{k_2} |
M_{1}\cdots M_{p}   m'_{1}m'_{2} \right \rangle \left \langle  m'_{1}m'_{2}|\widehat{J} ^{2}| m''_{1}m''_{2}\right \rangle  \left \langle M_{1}\cdots M_{p} m''_{1}m''_{2} | I^{k_2}  \right\rangle,
\end{eqnarray}
where $i=1,2$ and $l = 1,\cdots p$.  Although this equation looks
complicated, as there exist $k_2$, $ M_{1},\cdots M_{p} $, $ m''_{1}$, $m''_{2}$,
such that $\left \langle I^{k_2}|  M_{1}\cdots M_{p} m''_{1}m''_{2}
\right \rangle \neq 0$, Eq.(\ref{eq:eps18}) is nothing else but
\begin{eqnarray}\label{eq:eps19}
  \left \langle m'''_{1} m'''_{2}
|\widehat{J}^{2}| m_{1}m_{2} \right
\rangle  &=& K \delta _{m_{1}m'''_{1}} \delta _{m_{2}m'''_{2}} \nonumber \\
 &=& K \left \langle m'''_{1} m'''_{2}| m_{1} m_{2} \right
 \rangle,
\end{eqnarray}
where $K \in C$ is a constant. This means that
$\widehat{J}^{2}$ has only one eigenvalue $K$. However, when
$j_{1} \geq \frac{1}{2}$ and $j_{2} \geq  \frac{1}{2}$,
$\widehat{J}^{2}$ has at least two different eigenvalues
$(j_{1}+j_{2})(j_{1}+j_{2}+1)$ and $\left | j_{1} -j_{2} \right
|(\left |j_{1}-j_{2}  \right |+1)$. Therefore, our starting assumption is not true and the orthogonal relation as shown in Eq.(\ref{eq:eps8}) does not exist.

\centerline{\rule{80mm}{0.1pt}}


\begin{thebibliography}{99}
\vspace{3mm}

        \bibitem{Maldacena:2001kr}
        J.~M.~Maldacena,
        ``Eternal black holes in anti-de Sitter,''
        JHEP {\bf 0304}, 021 (2003)
        [hep-th/0106112].


        \bibitem{Ryu:2006bv}
        S.~Ryu and T.~Takayanagi,
        ``Holographic derivation of entanglement entropy from AdS/CFT,''
        Phys.\ Rev.\ Lett.\  {\bf 96}, 181602 (2006)
        [hep-th/0603001].

        \bibitem{VanRaamsdonk:2010pw}
        M.~Van Raamsdonk,
        ``Building up spacetime with quantum entanglement,''
        Gen.\ Rel.\ Grav.\  {\bf 42}, 2323 (2010)
        [Int.\ J.\ Mod.\ Phys.\ D {\bf 19}, 2429 (2010)]
        [arXiv:1005.3035 [hep-th]].

        \bibitem{Vidal:2007hda}
        G.~Vidal,
        ``Entanglement Renormalization,''
        Phys.\ Rev.\ Lett.\  {\bf 99}, no. 22, 220405 (2007)
        [cond-mat/0512165].

        \bibitem{Swingle:2009bg}
        B.~Swingle,
        ``Entanglement Renormalization and Holography,''
        Phys.\ Rev.\ D {\bf 86}, 065007 (2012)
        [arXiv:0905.1317 [cond-mat.str-el]].

        \bibitem{Swingle:2012wq}
        B.~Swingle,
        ``Constructing holographic spacetimes using entanglement renormalization,''
        arXiv:1209.3304 [hep-th].

        \bibitem{Nozaki:2012zj}
        M.~Nozaki, S.~Ryu and T.~Takayanagi,
        ``Holographic Geometry of Entanglement Renormalization in Quantum Field Theories,''
        JHEP {\bf 1210}, 193 (2012)
        [arXiv:1208.3469 [hep-th]].


        \bibitem{Qi:2013caa}
        X.~L.~Qi,
        ``Exact holographic mapping and emergent space-time geometry,''
        arXiv:1309.6282 [hep-th].

        \bibitem{Pastawski:2015qua}
        F.~Pastawski, B.~Yoshida, D.~Harlow and J.~Preskill,
        ``Holographic quantum error-correcting codes: Toy models for the bulk/boundary correspondence,''
        JHEP {\bf 1506}, 149 (2015)
        [arXiv:1503.06237 [hep-th]].

        \bibitem{Evenbly:2017htn}
        G.~Evenbly,
        ``Hyper-invariant tensor networks and holography,''
        Phys. Rev. Lett. {\bf 119}, 141602
        [arXiv:1704.04229 [cond-mat, physics:quant-ph]].


\bibitem{Ling:2018vza}
  Y.~Ling, Y.~Liu, Z.~Y.~Xian and Y.~Xiao,
  arXiv:1806.05007 [hep-th].

\bibitem{Ling:2018ajv}
  Y.~Ling, Y.~Liu, Z.~Y.~Xian and Y.~Xiao,
  arXiv:1807.10247 [hep-th].

\bibitem{Rovelli:1994ge}
  C.~Rovelli and L.~Smolin,
  Nucl.\ Phys.\ B {\bf 442}, 593 (1995)
  Erratum: [Nucl.\ Phys.\ B {\bf 456}, 753 (1995)]
  doi:10.1016/0550-3213(95)00150-Q, 10.1016/0550-3213(95)00550-5
  [gr-qc/9411005].

\bibitem{Rovelli:1995ac}
  C.~Rovelli and L.~Smolin,
  Phys.\ Rev.\ D {\bf 52}, 5743 (1995)
  doi:10.1103/PhysRevD.52.5743
  [gr-qc/9505006].

\bibitem{Orus:2014poa}
  R.~Orus,
  Eur.\ Phys.\ J.\ B {\bf 87}, 280 (2014)
  doi:10.1140/epjb/e2014-50502-9
  [arXiv:1407.6552 [cond-mat.str-el]].


\bibitem{Han:2016xmb}
  M.~Han and L.~Y.~Hung,
  Phys.\ Rev.\ D {\bf 95}, no. 2, 024011 (2017)
  doi:10.1103/PhysRevD.95.024011
  [arXiv:1610.02134 [hep-th]].

\bibitem{Chirco:2017vhs}
  G.~Chirco, D.~Oriti and M.~Zhang,
  Class.\ Quant.\ Grav.\  {\bf 35}, no. 11, 115011 (2018)
  doi:10.1088/1361-6382/aabf55
  [arXiv:1701.01383 [gr-qc]].

\bibitem{Chirco:2017xjb}
  G.~Chirco, F.~M.~Mele, D.~Oriti and P.~Vitale,
  Phys.\ Rev.\ D {\bf 97}, no. 4, 046015 (2018)
  doi:10.1103/PhysRevD.97.046015
  [arXiv:1703.05231 [gr-qc]].


\bibitem{Livine:2017fgq}
  E.~R.~Livine,
  Phys.\ Rev.\ D {\bf 97}, no. 2, 026009 (2018)
  doi:10.1103/PhysRevD.97.026009
  [arXiv:1709.08511 [gr-qc]].

\bibitem{Baytas:2018wjd}
  B.~Baytas, E.~Bianchi and N.~Yokomizo,
  Phys.\ Rev.\ D {\bf 98}, no. 2, 026001 (2018)
  doi:10.1103/PhysRevD.98.026001
  [arXiv:1805.05856 [gr-qc]].

\bibitem{LWX}
  Y. Ling, M. Wu, and Y. Xiao
  ``From quantum entanglement to quantum geometry'',
 to be published.

\end{thebibliography}
\end{document}